\documentclass[onecolumn,letterpaper,aps,prl,superscriptaddress,showpacs,floatfix]{revtex4-1}

\usepackage{graphicx}	
\usepackage{xspace}     
\usepackage{multirow}

\newcommand{\pt}{\mbox{$p_T$}\xspace}
\newcommand{\xt}{\mbox{$x_T$}\xspace}
\newcommand{\gevc}{\mbox{GeV/$c$}\xspace}

\newcommand{\raa}{\mbox{$R_{\rm AA}$}\xspace}

\newcommand{\Ncoll}{\mbox{$N_{\rm coll}$}\xspace}

\newcommand{\sqsn}{\mbox{$\sqrt{s_{_{\rm NN}}}$}\xspace}
\newcommand{\AuAu}{\mbox{Au$+$Au}\xspace}
\newcommand{\CuCu}{\mbox{Cu$+$Cu}\xspace}
\newcommand{\pp}{\mbox{$p$$+$$p$}\xspace}
\newcommand{\piz}{\mbox{$\pi^{0}$}\xspace}

\def\tab#1{Table~\ref{#1}}

\begin{document}


\title{Evolution of $\pi^{0}$ suppression in Au$+$Au collisions from
$\sqrt{s_{_{\rm NN}}}$ = 39 to 200 GeV
}

\newcommand{\abilene}{Abilene Christian University, Abilene, Texas 79699, USA}
\newcommand{\banaras}{Department of Physics, Banaras Hindu University, Varanasi 221005, India}
\newcommand{\barc}{Bhabha Atomic Research Centre, Bombay 400 085, India}
\newcommand{\baruch}{Baruch College, City University of New York, New York, New York, 10010 USA}
\newcommand{\bnlcoll}{Collider-Accelerator Department, Brookhaven National Laboratory, Upton, New York 11973-5000, USA}
\newcommand{\bnlphys}{Physics Department, Brookhaven National Laboratory, Upton, New York 11973-5000, USA}
\newcommand{\caucr}{University of California - Riverside, Riverside, California 92521, USA}
\newcommand{\charlesczech}{Charles University, Ovocn\'{y} trh 5, Praha 1, 116 36, Prague, Czech Republic}
\newcommand{\chonbuk}{Chonbuk National University, Jeonju, 561-756, Korea}
\newcommand{\cns}{Center for Nuclear Study, Graduate School of Science, University of Tokyo, 7-3-1 Hongo, Bunkyo, Tokyo 113-0033, Japan}
\newcommand{\colorado}{University of Colorado, Boulder, Colorado 80309, USA}
\newcommand{\columbia}{Columbia University, New York, New York 10027 and Nevis Laboratories, Irvington, New York 10533, USA}
\newcommand{\czechtech}{Czech Technical University, Zikova 4, 166 36 Prague 6, Czech Republic}
\newcommand{\dapnia}{Dapnia, CEA Saclay, F-91191, Gif-sur-Yvette, France}
\newcommand{\debrecen}{Debrecen University, H-4010 Debrecen, Egyetem t{\'e}r 1, Hungary}
\newcommand{\elte}{ELTE, E{\"o}tv{\"o}s Lor{\'a}nd University, H - 1117 Budapest, P{\'a}zm{\'a}ny P. s. 1/A, Hungary}
\newcommand{\ewha}{Ewha Womans University, Seoul 120-750, Korea}
\newcommand{\fsu}{Florida State University, Tallahassee, Florida 32306, USA}
\newcommand{\gsu}{Georgia State University, Atlanta, Georgia 30303, USA}
\newcommand{\hanyang}{Hanyang University, Seoul 133-792, Korea}
\newcommand{\hiroshima}{Hiroshima University, Kagamiyama, Higashi-Hiroshima 739-8526, Japan}
\newcommand{\ihepprot}{IHEP Protvino, State Research Center of Russian Federation, Institute for High Energy Physics, Protvino, 142281, Russia}
\newcommand{\illuiuc}{University of Illinois at Urbana-Champaign, Urbana, Illinois 61801, USA}
\newcommand{\inrras}{Institute for Nuclear Research of the Russian Academy of Sciences, prospekt 60-letiya Oktyabrya 7a, Moscow 117312, Russia}
\newcommand{\instpasczech}{Institute of Physics, Academy of Sciences of the Czech Republic, Na Slovance 2, 182 21 Prague 8, Czech Republic}
\newcommand{\isu}{Iowa State University, Ames, Iowa 50011, USA}
\newcommand{\jinrdubna}{Joint Institute for Nuclear Research, 141980 Dubna, Moscow Region, Russia}
\newcommand{\jyvaskyla}{Helsinki Institute of Physics and University of Jyv{\"a}skyl{\"a}, P.O.Box 35, FI-40014 Jyv{\"a}skyl{\"a}, Finland}
\newcommand{\kek}{KEK, High Energy Accelerator Research Organization, Tsukuba, Ibaraki 305-0801, Japan}
\newcommand{\korea}{Korea University, Seoul, 136-701, Korea}
\newcommand{\kurchatov}{Russian Research Center ``Kurchatov Institute", Moscow, 123098 Russia}
\newcommand{\kyoto}{Kyoto University, Kyoto 606-8502, Japan}
\newcommand{\labllr}{Laboratoire Leprince-Ringuet, Ecole Polytechnique, CNRS-IN2P3, Route de Saclay, F-91128, Palaiseau, France}
\newcommand{\lawllnl}{Lawrence Livermore National Laboratory, Livermore, California 94550, USA}
\newcommand{\losalamos}{Los Alamos National Laboratory, Los Alamos, New Mexico 87545, USA}
\newcommand{\lpc}{LPC, Universit{\'e} Blaise Pascal, CNRS-IN2P3, Clermont-Fd, 63177 Aubiere Cedex, France}
\newcommand{\lund}{Department of Physics, Lund University, Box 118, SE-221 00 Lund, Sweden}
\newcommand{\maryland}{University of Maryland, College Park, Maryland 20742, USA}
\newcommand{\mass}{Department of Physics, University of Massachusetts, Amherst, Massachusetts 01003-9337, USA }
\newcommand{\muhlenberg}{Muhlenberg College, Allentown, Pennsylvania 18104-5586, USA}
\newcommand{\myongji}{Myongji University, Yongin, Kyonggido 449-728, Korea}
\newcommand{\nagasaki}{Nagasaki Institute of Applied Science, Nagasaki-shi, Nagasaki 851-0193, Japan}
\newcommand{\newmex}{University of New Mexico, Albuquerque, New Mexico 87131, USA }
\newcommand{\nmsu}{New Mexico State University, Las Cruces, New Mexico 88003, USA}
\newcommand{\ohio}{Department of Physics and Astronomy, Ohio University, Athens, Ohio 45701, USA}
\newcommand{\ornl}{Oak Ridge National Laboratory, Oak Ridge, Tennessee 37831, USA}
\newcommand{\orsay}{IPN-Orsay, Universite Paris Sud, CNRS-IN2P3, BP1, F-91406, Orsay, France}
\newcommand{\pnpi}{PNPI, Petersburg Nuclear Physics Institute, Gatchina, Leningrad region, 188300, Russia}
\newcommand{\riken}{RIKEN Nishina Center for Accelerator-Based Science, Wako, Saitama 351-0198, Japan}
\newcommand{\rikjrbrc}{RIKEN BNL Research Center, Brookhaven National Laboratory, Upton, New York 11973-5000, USA}
\newcommand{\rikkyo}{Physics Department, Rikkyo University, 3-34-1 Nishi-Ikebukuro, Toshima, Tokyo 171-8501, Japan}
\newcommand{\saispbstu}{Saint Petersburg State Polytechnic University, St. Petersburg, 195251 Russia}
\newcommand{\saopaulo}{Universidade de S{\~a}o Paulo, Instituto de F\'{\i}sica, Caixa Postal 66318, S{\~a}o Paulo CEP05315-970, Brazil}
\newcommand{\seoulnat}{Department of Physics and Astronomy, Seoul National University, Seoul, Korea}
\newcommand{\stonybrkc}{Chemistry Department, Stony Brook University, SUNY, Stony Brook, New York 11794-3400, USA}
\newcommand{\stonycrkp}{Department of Physics and Astronomy, Stony Brook University, SUNY, Stony Brook, New York 11794-3400, USA}
\newcommand{\tenn}{University of Tennessee, Knoxville, Tennessee 37996, USA}
\newcommand{\titech}{Department of Physics, Tokyo Institute of Technology, Oh-okayama, Meguro, Tokyo 152-8551, Japan}
\newcommand{\tsukuba}{Institute of Physics, University of Tsukuba, Tsukuba, Ibaraki 305, Japan}
\newcommand{\vandy}{Vanderbilt University, Nashville, Tennessee 37235, USA}
\newcommand{\weizmann}{Weizmann Institute, Rehovot 76100, Israel}
\newcommand{\wigner}{Institute for Particle and Nuclear Physics, Wigner Research Centre for Physics, Hungarian Academy of Sciences (Wigner RCP, RMKI) H-1525 Budapest 114, POBox 49, Budapest, Hungary}
\newcommand{\yonsei}{Yonsei University, IPAP, Seoul 120-749, Korea}
\affiliation{\abilene}
\affiliation{\banaras}
\affiliation{\barc}
\affiliation{\baruch}
\affiliation{\bnlcoll}
\affiliation{\bnlphys}
\affiliation{\caucr}
\affiliation{\charlesczech}
\affiliation{\chonbuk}
\affiliation{\cns}
\affiliation{\colorado}
\affiliation{\columbia}
\affiliation{\czechtech}
\affiliation{\dapnia}
\affiliation{\debrecen}
\affiliation{\elte}
\affiliation{\ewha}
\affiliation{\fsu}
\affiliation{\gsu}
\affiliation{\hanyang}
\affiliation{\hiroshima}
\affiliation{\ihepprot}
\affiliation{\illuiuc}
\affiliation{\inrras}
\affiliation{\instpasczech}
\affiliation{\isu}
\affiliation{\jinrdubna}
\affiliation{\jyvaskyla}
\affiliation{\kek}
\affiliation{\korea}
\affiliation{\kurchatov}
\affiliation{\kyoto}
\affiliation{\labllr}
\affiliation{\lawllnl}
\affiliation{\losalamos}
\affiliation{\lpc}
\affiliation{\lund}
\affiliation{\maryland}
\affiliation{\mass}
\affiliation{\muhlenberg}
\affiliation{\myongji}
\affiliation{\nagasaki}
\affiliation{\newmex}
\affiliation{\nmsu}
\affiliation{\ohio}
\affiliation{\ornl}
\affiliation{\orsay}
\affiliation{\pnpi}
\affiliation{\riken}
\affiliation{\rikjrbrc}
\affiliation{\rikkyo}
\affiliation{\saispbstu}
\affiliation{\saopaulo}
\affiliation{\seoulnat}
\affiliation{\stonybrkc}
\affiliation{\stonycrkp}
\affiliation{\tenn}
\affiliation{\titech}
\affiliation{\tsukuba}
\affiliation{\vandy}
\affiliation{\weizmann}
\affiliation{\wigner}
\affiliation{\yonsei}
\author{A.~Adare} \affiliation{\colorado}
\author{S.~Afanasiev} \affiliation{\jinrdubna}
\author{C.~Aidala} \affiliation{\losalamos}
\author{N.N.~Ajitanand} \affiliation{\stonybrkc}
\author{Y.~Akiba} \affiliation{\riken} \affiliation{\rikjrbrc}
\author{R.~Akimoto} \affiliation{\cns}
\author{H.~Al-Ta'ani} \affiliation{\nmsu}
\author{J.~Alexander} \affiliation{\stonybrkc}
\author{A.~Angerami} \affiliation{\columbia}
\author{K.~Aoki} \affiliation{\riken}
% ...
\collaboration{PHENIX Collaboration} \noaffiliation

\date{\today}

\begin{abstract}

Neutral-pion, $\pi^{0}$, spectra were measured at midrapidity 
($|y|<0.35$) in Au$+$Au collisions at $\sqrt{s_{_{\rm NN}}}$ = 39 
and 62.4\,GeV and compared to earlier measurements at 200\,GeV in 
transverse-momentum range of $1 < p_T < 10$\,GeV/$c$.  
The high-$p_T$ tail is well described by a power law in all cases and 
the powers decrease significantly with decreasing center-of-mass 
energy.  
The change of powers is very similar to that observed in 
the corresponding spectra for $p$+$p$ collisions. 
The nuclear-modification factors ($R_{\rm AA}$) show significant
suppression, with a distinct energy, centrality, and $p_T$ 
dependence.
Above $p_T$= 7~Gev/$c$, $R_{\rm AA}$ is similar for 
$\sqrt{s_{_{\rm NN}}}$ = 62.4 and 200\,GeV at all centralities.
Perturbative-quantum-chromodynamics calculations that describe 
$R_{\rm AA}$ well at 200\,GeV fail to describe the 39\,GeV data, 
raising the possibility that for the same $p_T$ region, the relative 
importance of initial-state effects and soft processes increases 
at lower energies. 
The $p_T$ range where $\pi^0$ spectra in central Au$+$Au
collisions have the same power as in $p$+$p$ collisions is 
$\approx$~5 and 7 GeV/$c$ for $\sqrt{s_{_{\rm NN}}}$ = 200 and 62.4 ~GeV,
respectively.
For the $\sqrt{s_{_{\rm NN}}}$ = 39~GeV data, it is not clear whether 
such a region is reached, and the $x_T$-dependence of the $x_T$-scaling 
power-law exponent is very different from that observed in 
the $\sqrt{s_{_{\rm NN}}}$ = 62 and 200~GeV data, providing further 
evidence that initial-state effects 
and soft processes mask the in-medium suppression of hard-scattered 
partons to higher $p_T$ as the collision energy decreases.

\end{abstract}

	
\maketitle



Large transverse-momentum (\pt) particles produced in high-energy 
nucleus-nucleus (AB) collisions play a crucial role in studying 
the properties of the medium created in relativistic heavy-ion 
collisions.  Most hadrons at sufficiently high \pt are 
fragmentation products of hard-scattered partons and their 
production rate in vacuum, as measured in \pp collisions, is well 
described by perturbative quantum chromodynamics 
(pQCD)~\cite{ppg087}.  In the absence of any nuclear effects the 
production rate in relativistic heavy-ion collisions in the pQCD 
regime, i.e. at sufficiently high \pt, would scale with the 
increased probability that a hard scattering occurs, due to the 
large number of nucleons. 
This probability is characterized by the 
nuclear-overlap function $T_{\rm AB}$~\cite{glauber}.
However, such scaling has been violated to various degrees 
depending both on collision energy, \sqsn, and hadron \pt. At 
lower collision energies, the hadron yield is enhanced above the 
expected scaling.  This was first observed in $p$+A and this 
enhancement is generally attributed to multiple soft scattering 
(``Cronin effect''\cite{cronin}), and is presumed to occur in 
ion-ion collisions as well.  Initial parton-distribution 
functions in nuclei (nPDF) are different from those in 
protons~\cite{eskola2010}.

Finally, if a dense, colored medium is formed in the AB 
collision, the hard-scattered parton may traverse some of it, 
losing energy in the process.  Therefore, the observed yield at a 
given (high) \pt will be lower than that expected from $T_{\rm 
AB}$ scaling, exhibiting ``suppression'' or ``jet quenching,'' 
described in terms of the nuclear-modification factor, \raa (see 
Eq. (1)). Alternatively, other studies divide the yields for 
heavy-ion collisions at one energy with those for the same 
colliding species at a lower energy Au$+$Au , rather than scaled 
$p$+$p$ reference data, to study energy and centrality 
scaling~\cite{phobos}.

One of the first discoveries at the Relativistic Heavy-Ion 
Collider (RHIC) was a very large hadron suppression at high \pt 
(above $\approx$3\,\gevc) in \sqsn = 130 and 200\,GeV Au$+$Au 
collisions~\cite{ppg003,Adler:2002xw,ppg014,ppg080}. This 
suppression was attributed to the dominance of parton energy loss 
in the medium, i.e. to final-state effects. To test this 
hypothesis, the same measurements were performed in $d$$+$Au 
collisions~\cite{ppg028}, where the formation of the hot, dense 
partonic medium is not expected, and initial-state effects (if 
any) prevail.  No suppression in $d$$+$Au data was observed leaving 
little (if any) room for the initial-state effects as the origin 
of the large jet quenching observed in \AuAu. Studies with the 
lighter Cu+Cu system at three energies (\sqsn = 22.4, 62.4 and 
200\,GeV~\cite{ppg084}) have revealed that at \sqsn= 22.4\,GeV 
mechanisms that enhance \raa ($>1$) dominate at all centralities.  
Note, however, that this data set had very limited \pt range 
(\pt$<$4\,\gevc). At 62.4\,GeV, jet quenching overwhelms any 
enhancement and leads to a suppression (\raa$<1$) in more central 
collisions.

The low-energy scan at RHIC provides an opportunity to study the 
transition from enhancement (\raa$>1$) to suppression (\raa$<1$) 
and the evolution of \raa with collision energy, centrality and 
\pt.  The results put constraints on energy-loss models 
(see~\cite{brick} and references therein).
Here, we present new measurements by the PHENIX experiment at 
RHIC of \piz invariant yields and \raa in \AuAu collisions at 
\sqsn = 39 and 62.4\,GeV. The data were taken during the 2010 run 
and the \pt limits (statistics) were 8\,\gevc and 10\,\gevc, 
respectively. Reference \pp-collision data for \sqsn = 62.4\,GeV 
were taken in the same experiment in the 2006 run~\cite{ppg087}, 
while for \sqsn = 39\,GeV, data measured in the FERMILAB 
experiment E706 were used~\cite{e706}.

Neutral pions were measured on a statistical basis via their 
$\pi^0 \rightarrow \gamma\gamma$ decay branch with the 
electromagnetic calorimeter (EMCal)~\cite{nimemc}. The EMCal 
comprises two calorimeter types: 6 sectors of lead scintillator 
sampling calorimeter (PbSc) and 2 sectors of lead glass 
\v{C}erenkov calorimeter (PbGl). Each sector is located $\approx 
5$\,m from the beamline and subtends $|\eta| < 0.35$ in 
pseudorapidity and $\Delta\phi$ = 22.5$^\circ$ in azimuth. This 
Letter presents results obtained with the PbSc sectors only. The 
segmentation of the PbSc ($\Delta\eta\times\Delta\phi = 
0.01\times0.01$) ensures that the two photons from the $\pi^0 
\rightarrow \gamma\gamma$ decays are very well resolved up to \pt 
$<$ 12\,\gevc, i.e. across the entire \pt range of this 
measurement.

The results are based on data sets of $\,3.5\cdot 10^{8}\,$ and 
$\,7.0\cdot 10^{8}\,$ minimum-bias \AuAu events at 39 and 
62.4\,GeV, respectively. The minimum-bias (MB) trigger for both 
\sqsn = 39 and 62.4\,GeV was provided by the Beam-Beam-Counters 
(BBC)~\cite{nimbbc}, located close to the beam axis in both 
directions and covering $3.0\leq |\eta| \leq 3.9$.  In order to 
reduce background at least two hits were required in both BBCs. 
This condition selects $\sim86$\% of the total inelastic cross 
section. The centrality selection in \AuAu collisions at both 
energies was based on the charged signal sum of the BBCs, which 
is proportional to the charged particle multiplicity. For each 
centrality the average number of binary collisions ($\left<N_{\rm 
coll}\right>$) and the number of participants ($\left<N_{\rm 
part}\right>$) were calculated using a Glauber 
model~\cite{glauber} based Monte Carlo code.

 \begin{table}[thb]
\caption{Sources of systematic uncertainties and their relative 
effect [in \%] on the invariant yields for \sqsn = 39~GeV (62.4~GeV).
}
  \begin{ruledtabular}\begin{tabular}{rccccc}
      \pt: & \multicolumn{2}{c}{\,2\,\gevc\,} & 
\multicolumn{2}{c}{\,5\,\gevc\,} & Type\\
    \hline
      Yield extraction	& 3\% & (3\%)  & 3\% & (3\%) & A\\
      PID efficiency	& 4.5\% & (4.5\%)  & 4.5\% & (4.5\%) & B\\
      Energy scale	& 10.5\% & (8.0\%)  & 14.5\% & (10.0\%) & B \\
      Acceptance		& 2\% & (2\%) & 2\% & (2\%) & B \\
      Conversion		& 4\% & (4\%) & 4\% & (4\%) & B \\
      Off vertex		& 1.5\% & (1.5\%)& 1.5\% & (1.5\%) & C\\
\\
      Total for \piz yields & 12.7\% & (10.7\%)  & 16.2\% & (12.3\%) \\
  \end{tabular}\end{ruledtabular}
  \label{tab:syst}
 \end{table}
 
The PHENIX analysis of neutral pions is described in detail 
elsewhere~\cite{ppg080}. \tab{tab:syst} lists the sources of 
systematic uncertainties on the extracted-\piz invariant yields 
in this analysis. They can be divided into three different 
categories: (1) Type-A, \pt-uncorrelated; (2) Type-B, 
\pt-correlated, where the correlation may be an arbitrary smooth 
function; (3) Type-C, \pt-correlated, where all points move by 
the same fraction up or down.  The main sources of systematic 
uncertainties in the \piz measurement are the energy scale, yield 
extraction and particle-identification (PID) efficiency 
correction.
 
 \begin{figure}[htb]
 \includegraphics[width=0.85\linewidth]{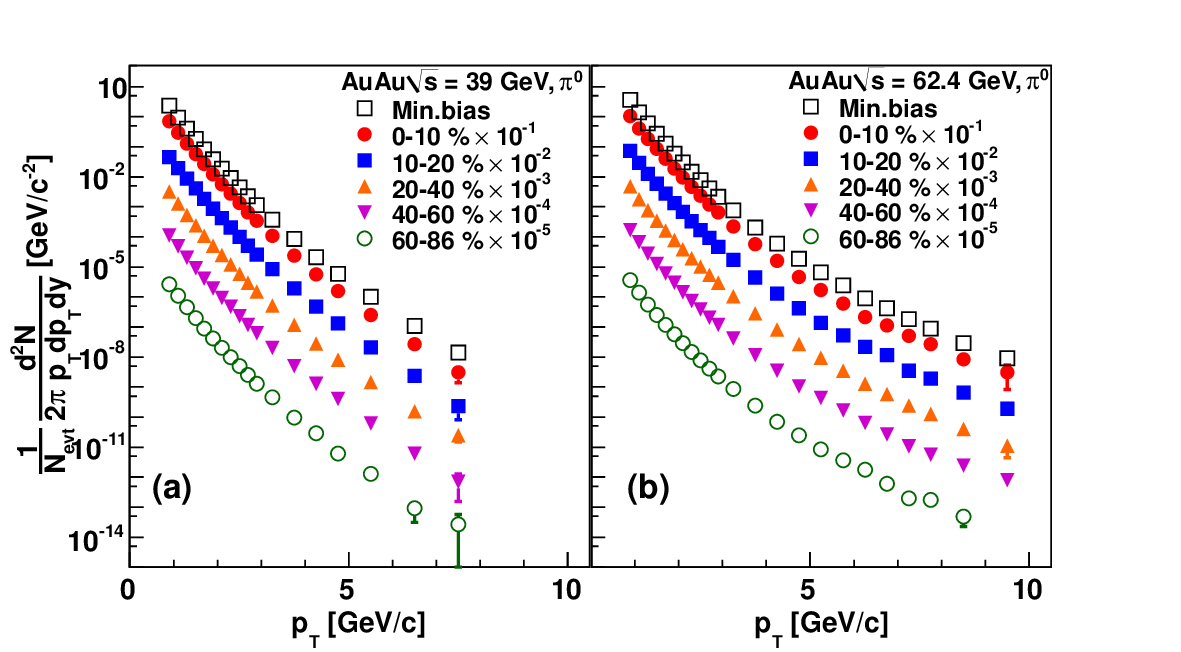}
    \caption{
    Invariant yields of \piz in \AuAu at \sqsn = 39\,GeV (a)
   and 62.4\,GeV (b) in all centralities and minimum
   bias. Only statistical uncertainties are shown.
   }
   \label{fig:invyield}
\end{figure}

Figure~\ref{fig:invyield} shows the invariant yields of the 
\piz{s} for all centralities and also in minimum-bias collisions. 
From fitting the \sqsn = 39 and 62.4\,GeV minimum-bias spectra 
with a power-law function ($\propto p_T^{n}$) for \pt $>$ 
4\,\gevc, we obtained powers $n_{39} = -12.07~\pm~0.18$ and 
$n_{62.4} = -10.60~\pm~0.09 $, respectively, significantly 
steeper than at \sqsn=200\,GeV, where $n_{200} = -8.06~\pm~0.08$ 
for MB collisions~\cite{ppg080}. The slopes of the corresponding 
\pp-collision spectra are somewhat different, but comparable, 
$n_{39}^{pp}=-12.02~\pm~0.31$, $n_{62.4}^{pp}=-9.82 \pm 0.18$ and 
$n_{200}^{pp}=-8.22 \pm 0.09$, respectively.

Nuclear effects on the \piz production are quantified using the
nuclear modification factor
  \begin{equation}
  R_{\rm AA} (p_T) =  \frac{(1/N_{\rm AA}^{\rm evt}) {\rm d}^2N_{\rm AA}^{\pi^0}/{\rm d}p_T{\rm dy}}{\left<T_{\rm AB}\right>\times{\rm d}^2\sigma_{\rm pp}^{\pi^0}/{\rm d}p_T{\rm dy}},
  \end{equation}
where $\sigma_{pp}^{\pi^0}$ is the production cross section of 
\piz in \pp collisions, and $\left<T_{\rm AB}\right> = 
\left<N_{\rm coll}\right>/\sigma_{pp}^{\rm inel}$ is the 
nuclear-overlap function averaged over the range of impact 
parameters contributing to the given centrality class according to 
the Glauber model. Thus \raa compares the yield observed in A$+$A 
collisions to the yield expected from the superposition of \Ncoll 
independent \pp interactions.  In the absence of nuclear effects, 
\raa should be equal to unity.  However, \raa$\approx 1$ does not 
necessarily imply the absence of suppression, it may also 
indicate a balance between enhancing and depleting mechanisms.

In order to calculate \raa, a reference \pt distribution in \pp 
collisions is needed.  Preferably this is measured with the same 
detector, in which case many systematic uncertainties cancel in 
the ratio.  The PHENIX experiment has measured the \piz cross 
section in \pp collisions at \sqsn = 62.4 \,GeV~\cite{ppg087} but 
only up to \pt = 7\,\gevc while the current Au$+$Au measurement 
reaches up to 10\,\gevc.  Hence the \pp data were fitted with a 
power-law function between $4.5 < p_T < 7$\,\gevc and then 
extrapolated.  The systematic uncertainty resulting from this 
extrapolation reaches 20\% at 10\,\gevc, estimated from a series 
of fits, where each time one or more randomly selected points are 
omitted and the remaining points are re-fitted.

Because PHENIX has not measured the \pp spectrum of \piz at 
\sqsn = 39\,GeV, data from the Fermilab experiment 
E706~\cite{e706} were used.  However, the E706 acceptance 
($-1.0<|\eta|<0.5$) is different from that of PHENIX 
($|\eta|<0.35$), and since $dN/d\eta$ is not flat and narrows 
for high-\pt particles, a \pt-dependent correction was applied 
to the E706 data.  This correction factor was determined from a 
{\sc pythia} simulation by means of the ratio of yields 
(normalized per unit rapidity) when calculated from the observed 
yield in the PHENIX and E706 acceptance windows.  The systematic 
uncertainty of the correction is 1--2\% at 3\,\gevc but reaches 
20\% at 8\,\gevc.

\begin{figure}[tbh]
   \includegraphics[width=0.75\linewidth]{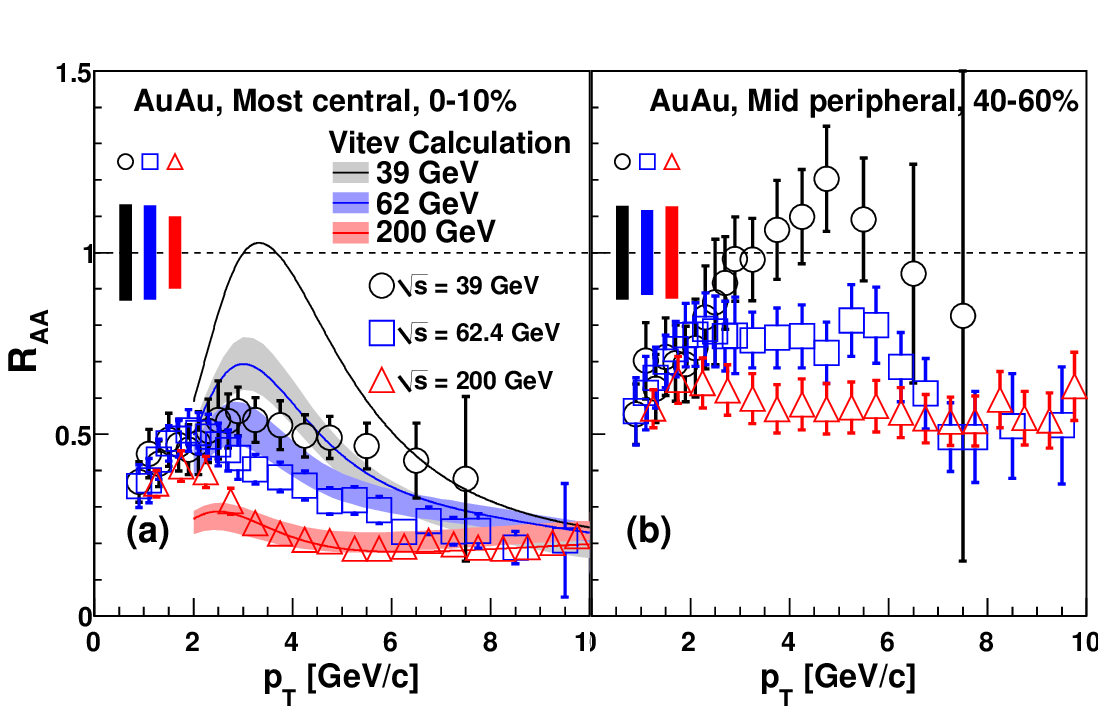}
   \caption{
   Nuclear modification factor (\raa) of \piz in \AuAu
   collisions in most central 0--10\% (a) and mid-peripheral
   40--60\% (b).  Error bars are the quadratic sum of statistical
   and \pt-correlated systematic uncertainties (including systematic
   uncertainties from the \pp-collision reference).
   Boxes around 1 are the quadratic sum of the C-type
   uncertainties combined with the \Ncoll uncertainties.  These are
   fully correlated between different energies.
      Also shown for central collisions are pQCD 
      calculations~\protect\cite{privatevitev}
      with Cronin effect as implemented in~\protect\cite{vitev} (solid lines) and with 
      Cronin effect 
      corresponding to 30\% larger initial-state parton mean free paths for all three 
      energies (bands).
}
   \label{fig:raa}
   \end{figure}

Figure~\ref{fig:raa} shows the nuclear modification factor of 
\piz{s} measured in \AuAu collisions at \sqsn = 39, 62.4 and 
200\,GeV (data from~\cite{ppg080}) as a function of \pt for most 
central collisions (a) and 40--60\% centrality (b).  In the most 
central collisions (0--10\%) there is a significant suppression 
for all three energies, while in mid-peripheral collisions 
(40--60\%) at \sqsn=39\,GeV, \raa is consistent with unity above 
\pt $>$ 3\,\gevc.

Figure~\ref{fig:raa} also shows pQCD 
calculations~\cite{vitev,privatevitev} for 0--10\% centrality. 
The solid curves are obtained with a parametrization of 
initial-state multiple scattering~\cite{vitev} that 
overestimates the Cronin effect.  At high \pt, the theoretical 
result is compatible with the 200\,GeV \AuAu data (and also 
200\,GeV \CuCu data~\cite{ppg084}). Neither the 62.4, nor the 
39\,GeV data are consistent with the predictions.  The only 
qualitative agreement is that the turnover point of the \raa 
curves moves to higher \pt with lower collision energy as 
observed in the data. The bands are calculated within the same 
framework but with 30\% larger initial-state parton mean free 
paths and the energy loss varied by $\pm$10\%. The Cronin effect 
is then compatible with lower energy p+A data and earlier 
calculations~\cite{Vitev:2004gn}. The 200\,GeV data are still 
well described, the 62.4\,GeV data are consistent within 
uncertainties, but the 39\,GeV \raa, particularly the shape, is 
inconsistent with the corresponding band.

\begin{figure}[htb]
   \includegraphics[width=0.75\linewidth]{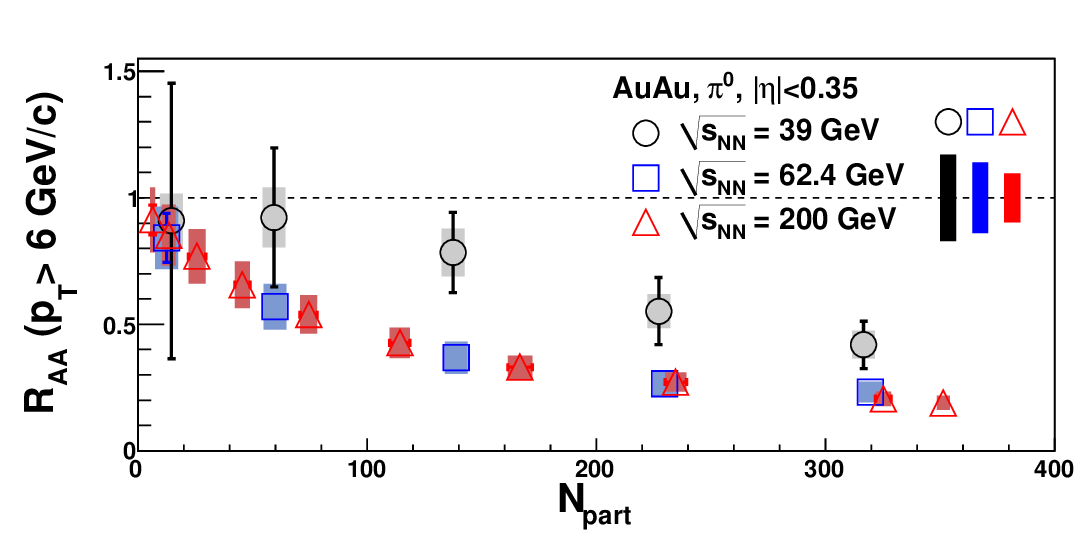}
      \caption{
      Nuclear modification factor averaged 
      for $p_T>6$\,\gevc.  Uncertainties are shown as error bars
      (statistical), boxes (sum of \pt-uncorrelated and \Ncoll), boxes
      around one (Type B and C and uncertainties from the \pp-collision reference).
      }
   \label{fig:intraa}
\end{figure}

Coupled with the observations that the slopes at high \pt become 
much steeper, but the bulk properties (like elliptic flow, energy 
density, apparent temperature) change only slowly in the 
collision-energy range in question, it is quite conceivable that 
hard scattering as a source of particles at a given \pt becomes 
completely dominant only at higher transverse momentum, i.e. jet 
quenching will be ``masked" up to higher \pt. Note that while the 
shapes at lower \pt are different, at \pt $>\approx$ 7\,\gevc 
\raa is essentially the same for the 62.4 and 200 GeV data, 
irrespective of centrality (see also Fig.~\ref{fig:intraa}).  
The simultaneous description of results spanning such a wide 
range in \sqsn is a challenge for energy-loss models that must 
incorporate multiple effects beyond radiative energy loss -- 
effects that may each have a different dependence on \sqsn. 
  
\begin{figure}[htp]
   \includegraphics[width=0.8\linewidth]{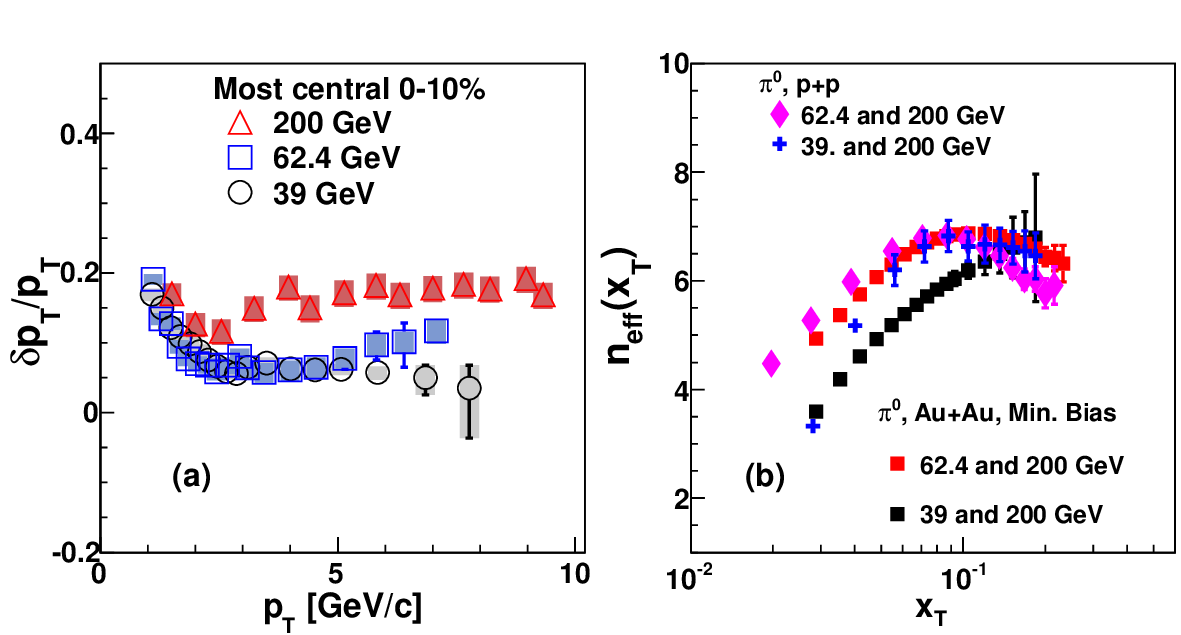}
      \caption{(Color online)
      (a) Fractional momentum shift $\delta p_T / p_T$ between \AuAu
      and $T_{\rm AB}$-scaled \pp data as a function of the \AuAu \pt.
      (b) Power $n_{eff}$ of $x_T$-scaling for \pp and \AuAu (minimum bias) 
      at various collision energies. }
   \label{fig:xtratio}
\end{figure}

Figure~\ref{fig:intraa} shows \pt-averaged \raa as a function of 
the number of participants. The averaging was done above 
$p_T>$6\,\gevc.  Our first observation is that \raa decreases 
with increasing centrality even for the lowest-energy system.  
Similarly, as already discussed in the context of 
Fig.~\ref{fig:raa}, at high enough \pt the suppression is the 
same at 62.4 and 200\,GeV, at all centralities.  This is 
remarkable because the power $n$ of the fit to the spectra 
changes approximately by two units from 200 to 62.4\,GeV, so the 
average momentum loss of the partons also has to be different in 
order to compensate the effect of the changing slope. The average 
momentum loss is usually defined by the fractional momentum shift 
$\delta p_T / p_T$ between the corresponding \AuAu and $T_{\rm 
AB}$-scaled \pp spectra as follows. Since the power-law tails of 
the \pp and \AuAu spectra are similar, they can be fitted 
simultaneously with the same function and same power $n$
  \begin{equation}
  f(p_T) =  \frac{A}{(p_T (1 +  \delta p_T / p_T))^n}, 
  \end{equation}
with $\delta p_T$ being the horizontal shift between the scaled 
\pp and the \AuAu spectra.  In panel (a) of 
Fig.~\ref{fig:xtratio}, the observed fractional momentum shifts 
are shown for central collisions, as a function of the \AuAu \pt.
This shows partons in 200 GeV collisions suffer the largest 
average momentum loss compared to the lower energies.

Inclusive single-particle spectra at sufficiently high \pt and 
collision energy were predicted to exhibit scaling with the 
variable $x_T=2p_T/\sqrt{s}$ such that the production cross 
section can be written in a form~\cite{xtscaling,brodsky}
\begin{equation}
  E\frac{d^3\sigma}{dp^3} =  
  \frac{1}{\sqrt{s}^{n(x_T,\sqrt{s})}} G(x_T),
\end{equation}
where $G(x_T)$ is a universal function and $n(x_T,\sqrt{s})$ 
characterizes the specific process~\cite{brodsky}.  The scaling 
power $n_{\rm eff}(x_T)$ between any pair of \sqsn energies is 
then calculated as
\begin{equation}
  n_{\rm eff}(x_T) = \frac{\log{(Yield(x_T,\sqrt{s_1})/Yield(x_T,\sqrt{s_2}))}}
  {\log{(\sqrt{s_2}/\sqrt{s_1})}}.
\end{equation}

In panel (b) of Fig.~\ref{fig:xtratio}, $n_{\rm eff}(x_T)$ is 
shown when comparing invariant-\piz yields in \pp and \AuAu 
collisions at different energies.  Both the shape and the 
magnitude of $n_{\rm eff}(x_T)$ is similar for the 62.4/200\,GeV 
\pp and \AuAu as well as for the 39/200\,GeV \pp data. The rise 
of $n_{\rm eff}(x_T)$ at lower \xt can be attributed to the 
dominance of soft processes~\cite{lowxt}, while at higher \xt 
they deviate strongly from leading-twist scaling 
predictions~\cite{leadingtwist,brodsky}. 
However, for 39/200\,GeV, we
observe a significant difference for $n_{\rm eff}(x_T)$ 
for \pp compared to \AuAu collisions.
It may not even reach its 
maximum in the measured-\xt range, and its constant rise is 
similar to the rise observed in the low-\xt (soft) region of the 
other data shown.  One possible explanation could be that while 
present, hard scattering is still not the overwhelming source of 
high-\pt \piz{s} in the currently available \pt range in 39\,GeV 
\AuAu collisions.

In summary, the \piz \pt spectra were measured in \AuAu 
collisions at two different energies, \sqsn = 39 and 62.4\,GeV, 
and compared to the earlier result for \sqsn = 200\,GeV. In all 
cases the high \pt part of the invariant yields can be well 
described with a single power-law function.  The powers decrease 
considerably at lower \sqsn, and since the soft processes change 
only slowly with collision energy, jet quenching might be 
``masked'' up to higher transverse momenta. The high-\pt \piz 
yields in \AuAu at 62.4\,GeV are suppressed, and above 
\pt~$>$~6\,\gevc the data points are comparable with the 200\,GeV 
results at all centralities. The \piz yields in \AuAu at 39\,GeV 
are suppressed in the most central collisions, but no suppression 
is apparent in more peripheral collisions. At lower energies, a 
decreasing momentum shift compensates for the steeper slopes at 
high \pt, making the \raa's comparable, in fact, identical in the 
case of 62.4 and 200\,GeV.  When related to 200\,GeV, $n_{\rm 
eff}(x_T)$ is similar for 62.4 and 39\,GeV \pp and 62.4\,GeV 
\AuAu, but very different for the 39\,GeV \AuAu data. 
Measurements of flow, multiplicity, and other quantities
indicate that the soft processes vary slowly with \sqsn~\cite{ppg019}.
Also the Cronin effect, which counteracts suppression, is
actually increasing with decreasing energy~\cite{Adil:2004cn}.  This coupled
with the rapid decrease of the high-\pt slope with decreasing
energy, masks the in-medium suppression of hard-scattered
partons up to higher \pt.



\begin{acknowledgments}  


We thank the staff of the Collider-Accelerator and Physics
Departments at Brookhaven National Laboratory and the staff of
the other PHENIX participating institutions for their vital
contributions.  
We also thank Ivan Vitev for valuable discussions and
helpful suggestions for changes to the text.
We acknowledge support from the 
Office of Nuclear Physics in the
Office of Science of the Department of Energy, the
National Science Foundation, Abilene Christian University
Research Council, Research Foundation of SUNY, and Dean of the
College of Arts and Sciences, Vanderbilt University (U.S.A),
Ministry of Education, Culture, Sports, Science, and Technology
and the Japan Society for the Promotion of Science (Japan),
Conselho Nacional de Desenvolvimento Cient\'{\i}fico e
Tecnol{\'o}gico and Funda\c c{\~a}o de Amparo {\`a} Pesquisa do
Estado de S{\~a}o Paulo (Brazil),
Natural Science Foundation of China (P.~R.~China),
Ministry of Education, Youth and Sports (Czech Republic),
Jyv{\"a}skyl{\"a} University (Finland),
Centre National de la Recherche Scientifique, Commissariat
{\`a} l'{\'E}nergie Atomique, and Institut National de Physique
Nucl{\'e}aire et de Physique des Particules (France),
Bundesministerium f\"ur Bildung und Forschung, Deutscher
Akademischer Austausch Dienst, and Alexander von Humboldt Stiftung (Germany),
Hungarian National Science Fund, OTKA (Hungary), 
Department of Atomic Energy and Department of Science and Technology (India), 
Israel Science Foundation (Israel), 
National Research Foundation and WCU program of the 
Ministry Education Science and Technology (Korea),
Ministry of Education and Science, Russian Academy of Sciences,
Federal Agency of Atomic Energy (Russia),
VR and Wallenberg Foundation (Sweden), 
the U.S. Civilian Research and Development Foundation for the
Independent States of the Former Soviet Union, 
the Hungarian American Enterprise Scholarship Fund,
and the US-Israel Binational Science Foundation.  

\end{acknowledgments}  



This letter has a published Erratum (see Ref.~\cite{ppg138erratum}) with
the text:  ``We previously reported neutral-pion ($\pi^{0}$) spectra
measured at midrapidity ($|y|<0.35$) in Au$+$Au collisions at
$\sqrt{s_{_{NN}}}=39$ and 62.4~GeV, which were compared to
earlier measurements at 200~GeV in the transverse-momentum
range of $1<p_T<10$~GeV/$c$.  Subsequently, we discovered that
in Fig.~1 the numbers on the y-axis (invariant yields) are
wrong.  The data points, discussion, and conclusions are
unchanged.  This Erratum simply shows Fig.~1 with correct
numbers on the ordinate." (Fig. 1 is replaced above).

\end{document}